\documentclass[prl,twocolumn,showpacs,preprintnumbers,amsmath,amssymb]{revtex4}
\usepackage{graphicx,bm}
\def\v#1{\bm #1 }
\def\dfrac#1#2{{\displaystyle\frac{#1}{#2}}}
\def\H{{\cal H}}

\def\ln{\mbox{ln}}

\def\JA{J_{\mbox{A}}}
\def\JB{J_{\mbox{B}}}
\def\XA{X_{\mbox{A}}}
\def\XB{X_{\mbox{B}}}

\def\const{{\mbox{const.}}}

\begin{document}
\title{New Universality Class in the S=1/2 Fibonacci Heisenberg Chains}

\author{Kazuo Hida}
 \affiliation{Department of Physics, Faculty of Science,\\ Saitama University, Saitama, Saitama 338-8570, JAPAN
}
 \email{hida@phy.saitama-u.ac.jp}
\date{\today}

\begin{abstract}
Low energy properties of the $S=1/2$ antiferromagnetic Heisenberg chains with Fibonacci exchange modulation are studied using the real space renormalization group method for strong exchange modulation. Using the analytical solution of the recursion equation, the true asymptotic behavoir is revealed, which was veiled by the finite size effect in the previous numerical works. It is found that the ground state of this model belongs to a new universality class with logarithmically divergent dynamical exponent which is neither like Fibonacci XY chains nor like XY chains with relevant aperiodicity.
\end{abstract}


\sloppy
\maketitle
The magnetism of quasiperiodic systems has been the subject of continual studies since the discovery of quasicrystals in 1984\cite{shecht}. This problem has been attracting renewed interest after the synthesis of magnetic quasicrystals with well-localized magnetic moments\cite{sato1}. The artificial formation of one and two dimensional quasiperiodic structure is also coming into  the scope of experimental physics\cite{dot1,ledieu} thanks to the recent progress of nanotechnology and surface engineering. Possiblly motivated by these experimental progress, the theoretical investigation of the quantum magnetism in one and two dimensional quasiperiodic systems are started by many authors\cite{kh1,kh2,vidal1,acg,wjh,jag}.   

The  $S=1/2$ Fibonacci XY chain, which  is mapped onto the free fermion chain, has been studied extensively by  Kohmoto and coworkers\cite{kkt1}  by means of the exact renormalization group method from the early days of quasicrystal physics. It is shown that the ground state of the XY chain with Fibonacci exchange modulation is critical with finite non-universal dynamical exponents. This approach was extended to include other types of aperiodicity and anisotropy\cite{jh1}. It is clarified that the criticality of the Fibonacci XY chain emerges from the marginal nature of the Fibonacci and other precious mean aperiodicity in this model. For the relevant aperiodicity, more singular behavior with a divergent dynamical exponent is realized even for the XY chain\cite{jh1}.

On the other hand, the investigation of the $S=1/2$ Fibonacci Heisenberg chains started only in late 90's. The ground state of the uniform $S=1/2$ Heisenberg chain is exactly solved  by the famous Bethe ansatz method\cite{taka} and is known to be in the Luttinger liquid state with conformal invariance. This implies that the dynamical exponent $z$  is unity and the specific heat $C$ and susceptibility $\chi$ behave as $C \sim T$ and $\chi \sim \const$ at low temperatures. This exact solution is related to the transfer matrix of the 2-dimensional classical 8 vertex model which can be solved exactly.\cite{taka} On the quasiperiodic lattice also, some 2-dimensional classical models are known to have exact solution.\cite{kore}  However, the exact solution of the Fibonacci Heisenberg chain is not derived from these exactly solvable vertex models. Therefore we must resort to the renormalization group approach to clarify the reliable low energy asymptotic behavior. For weak Fibonacci modulation, Vidal and coworkers\cite{vidal1} have shown that the Fibonacci modulation is relevant on the basis of the weak modulation renormalization group calculation. The present author carried out the DMRG calculation and investigated the scaling properties of the low energy spectrum\cite{kh1,kh2}. In the present work, we employ the real space renormalization group (RSRG) method\cite{df1},  which is valid for the strong modulation, to elucidate the ground state properties of the  $S=1/2$ antiferromagnetic Fibonacci Heisenberg chains. Surprisingly, the finite size scaling formula, which fitted the DMRG data in ref. \cite{kh1} well, turned out to be the artifact of the finite size crossover effect. The true asymptotic behavior is first revealed by the exact solution of the recursion equation obtained in the present paper. It is also explained why the DMRG data are well fitted by the formula assumed in ref \cite{kh1} within the appropriate range of the system size.

Our Hamiltonian is given by,
\begin{equation}
\label{eq:ham}
\H = \sum_{i=1}^{N-1} J_{\alpha_i}\v{S}_{i}\v{S}_{i+1},\ \ \ (J_{\alpha_i} > 0,\ \alpha_i=A \ \mbox{or} \ B),
\end{equation}
where $\v{S}_{i}$'s are the spin 1/2 operators.  The exchange couplings $J_{\alpha_i}$ ($=\JA$ or $\JB$) follow the Fibonacci sequence generated by the substitution rule,
\begin{equation}
A \rightarrow AB, \ B \rightarrow A.
\label{subs}
\end{equation}

If one of the couplings $\JA$ or $\JB$ is much larger than the other, we can decimate the spins coupled via the stronger exchange coupling and calculate the effective interaction between the remaining spins by the perturbation method with respect to the weaker coupling\cite{df1}. This type of decimation scheme has been used to investigate the magnetization process of the Fibonacci Heisenberg chains\cite{acg}. Here we apply this scheme to find the fixed point which governs the ground state in the absence of magnetization.

\begin{figure}
\centerline{\includegraphics[height=20mm]{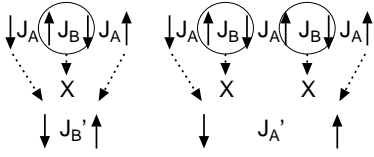}}
\caption{The decimation procedure for  $\JA << \JB$.}
\label{deci1}
\end{figure}
\begin{figure}
\centerline{\includegraphics[width=90mm]{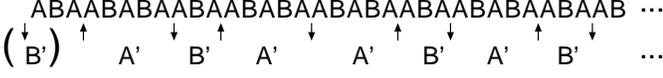}}
\caption{The RSRG scheme of the Fibonacci Heisenberg chain. The letters A and B correspond to the bonds and the up and down arrows to the spins which survive  decimation.  For $\JA >> \JB$ the leftmost  spin and bond in the parenthesis do not appear.}
\label{renorm}
\end{figure}
\begin{figure}
\centerline{\includegraphics[height=20mm]{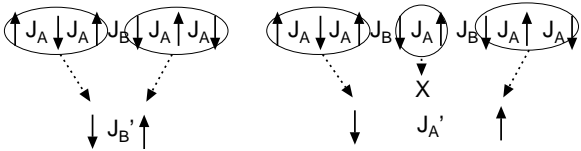}}
\caption{The decimation procedure for  $\JA >> \JB$.}
\label{deci2}
\end{figure}
In the present work, we concentrate on the case of strong modulation, max($\JA/\JB, \JB/\JA) >> 1$. For $\JA << \JB$, the spins connected by the $\JB$-bonds are decimated as shown in Fig. \ref{deci1}. The spin-1/2 degrees of freedom survive on the sites in the middle of the sequence AA. The two kinds of sequences of bonds are allowed between two alive spins, namely ABA and ABABA.  Between these two alive spins,  there exists one singlet pair in the former case while  two singlet pairs exist in the latter case. Therefore the effective coupling is weaker for the latter case. This decimation process  replaces the sequence ABABA by A' and ABA sandwiched by two As by B' resulting in the sequence B'A'B'A'A'B'A'B'A'... Except for B' at the leftmost position,  this sequence again gives the Fibonacci sequence as schematically shown in Fig. \ref{renorm}. As seen from the change of the number of the bonds by one step of decimation, this procedure essentially corresponds to a 3-step deflation. The rigorous proof will be reported in a separate paper\cite{kh3}.

In the case  $\JA >> \JB$, the decimation precesses are shown in Fig. \ref{deci2}. The three spins connected by successive A bonds form a doublet which can be described as a single spin with magnitude 1/2. The spins connected by the isolated $\JA$ bonds are decimated. Therefore this decimation process again corresponds to the replacement  ABABA $\rightarrow$ A' and ABA$\rightarrow$ B'. In this case, the resulting sequence is the exact Fibonacci sequence. After the first decimation the A' bond becomes weaker than the B' bonds. Therefore the decimation rule for the case $\JA << \JB$ applies for the further decimation procedure. 

The effective coupling can be calculated by the straightforward perturbation theory in weaker coupling. For the $n$-th iteration, we have,
\begin{eqnarray}
\JA^{(n+1)}&=&\frac{\JA^{(n)3}}{4\JB^{(n)2}}, \ \ \JB^{(n+1)}=\frac{\JA^{(n)2}}{2\JB^{(n)}}, 
\label{iter}
\end{eqnarray}
with
\begin{eqnarray}
\JA^{(1)}&=&\frac{\JA^{3}}{4\JB^{2}}, \ \  \JB^{(1)}=\frac{\JA^{2}}{2\JB}\ \ \mbox{for} \ \ \JB >> \JA , \\
\JA^{(1)}&=&\frac{2\JB^2}{9\JA}, \ \  \JB^{(1)}=\frac{4\JB}{9}\ \ \mbox{for} \ \ \JA >> \JB ,
\end{eqnarray}
where the variables with  $^{(n)}$ refer to the values after $n$-step iteration.

The ratio $\JA^{(n)}/\JB^{(n)}$ decreases under renormalization as,
\begin{eqnarray}
\frac{\JA^{(n+1)}}{\JB^{(n+1)}}&=&\frac{1}{2}\frac{\JA^{(n)}}{\JB^{(n)}} \ \ \ (n \geq  1).
\end{eqnarray}
This implies that the perturbation approximation becomes even more accurate as the renormalization proceeds. Therefore the aperiodicity is  relevant in consistency with the result of the weak modulation renormalization group method\cite{vidal1}. Taking the both results into account, we may safely expect that the ground state of the Fibonacci Heisenberg chain is governed by the strong modulation fixed point obtained in the present approach in the entire parameter range of $\JA/\JB \ne 1$.

The solution of the recursion equation (\ref{iter}) is given by,
\begin{eqnarray}
\left.
\begin{array}{l}
\JA^{(n)} =\JA\left(\dfrac{\JA}{\JB}\right)^{2n}2^{-n(n+1)}\\
\JB^{(n)} =\JB\left(\dfrac{\JA}{\JB}\right)^{2n}2^{-n^2} 
\end{array}\right\}\ \ \mbox{for} \ \ \JB >> \JA,
\end{eqnarray}
\begin{eqnarray}
\left.
\begin{array}{l}
\JA^{(n)} =\dfrac{8\JB}{9}\left(\dfrac{\JB}{\JA}\right)^{2n-1}2^{-n(n+1)}\\
\JB^{(n)} =\dfrac{8\JA}{9}\left(\dfrac{\JB}{\JA}\right)^{2n-1}2^{-n^2} 
\end{array}\right\}\ \ \mbox{for} \ \ \JB << \JA.
\end{eqnarray}

The length of the $3n$-th Fibonacci sequence is equal to the Fibonacci number $F_{3n}$ which grows as $\phi^{3n}$ for large $n$ where $\phi$ is the golden mean ($=\frac{1+\sqrt{5}}{2}$). Therefore the chain of length $N \sim \phi^{3n}$ reduces to a single pair of spins after $n$ decimation steps. This implies that the smallest energy scale $\Delta E$ for the finite Fibonacci chain with length $N$ scales as,
\begin{eqnarray}
\lefteqn{\Delta E \sim 2^{-n^2}\sim \exp\left({-(\ln N /3\ln \phi)^2\ln 2}\right)}\nonumber\\
&=& e^{-\kappa(\ln N)^2 }= N^{-\kappa\ln N }\ \ \mbox{with}\ \ \kappa\equiv \ln 2/(3\ln \phi)^2 
\label{true}
\end{eqnarray}
for large enough $N$, irrespective of the value of $\dfrac{\JA}{\JB}$. It should be noted that the dynamical exponent diverges logarithmically.

This size dependence implies that the number of the magnetic excited states with energies in the interval $\Delta E \sim \Delta E + d\Delta E$ the magnetic excitation $ND(\Delta E)d\Delta E$ scales as,
\begin{eqnarray}
\lefteqn{ND(\Delta E)d\Delta E}  \nonumber\\
&\sim& f\left(N\exp\left(-\sqrt{\frac{1}{\kappa}\ln \frac{1}{\Delta E}}\right)\right)d\left(N\exp\left(-\sqrt{\frac{1}{\kappa}\ln \frac{1}{\Delta E}}\right)\right) \nonumber\\
&\sim& Nf\left(N\exp\left(-\sqrt{\frac{1}{\kappa}\ln \frac{1}{\Delta E}}\right)\right)\nonumber\\
&\times&\frac{1}{2\kappa\Delta E \sqrt{\frac{1}{\kappa}\ln \frac{1}{\Delta E}}}\exp\left(-\sqrt{\frac{1}{\kappa}\ln \frac{1}{\Delta E}}\right)d\Delta E 
\end{eqnarray}
with a scaling function $f(x)$. Because the density of state per site $D(\Delta E)$ should be finite in the thermodynamic limit $N \rightarrow \infty$, the scaling function $f(x)$ tends to a finite value as $x \rightarrow \infty$. Therefore we find,
\begin{eqnarray}
\lefteqn{D(\Delta E)d\Delta E} \nonumber\\
&\sim& \frac{1}{2\kappa\Delta E \sqrt{\frac{1}{\kappa}\ln \frac{1}{\Delta E}}}\exp\left(-\sqrt{\frac{1}{\kappa}\ln \frac{1}{\Delta E}}\right)d\Delta E 
\end{eqnarray}
for large enough $N$. Accordingly, the low temperature magnetic specific heat $C$ should behave as,
\begin{eqnarray}
C &\sim& \frac{\partial}{\partial T}N\int_0^T \Delta E D(\Delta E) d\Delta E \sim NTD(T) \nonumber\\ 
 &\sim&  \frac{N}{2\kappa  \sqrt{\frac{1}{\kappa}\ln \frac{1}{T}}}\exp\left(-\sqrt{\frac{1}{\kappa}\ln \frac{1}{T}}\right).
 \label{spe}
 \end{eqnarray}
The magnetic susceptibility at temperature $T$ should be the Curie contribution from the spins alive at the energy scale $T$. The number $n_s(T)$ of  such spins is given by 
\begin{eqnarray}
n_s(T)&\sim& 2N\int_0^T D(\Delta E) d\Delta E ,
\end{eqnarray}
because two spins are excited by breaking a single effective bond with effective exchange energy less than $k_BT$. Therefore  the low temperature magnetic susceptibility $\chi$ behaves as,
\begin{eqnarray}
\chi(T)&\sim& \frac{2N}{4T}\int_0^T \frac{\exp\left(-\sqrt{\frac{1}{\kappa}\ln \frac{1}{\Delta E}}\right)}{2\kappa\Delta E \sqrt{\frac{1}{\kappa}\ln \frac{1}{\Delta E}}}d\Delta E \nonumber \\ 
&\sim& \frac{N\exp\left(-\sqrt{\frac{1}{\kappa}\ln \frac{1}{T}}\right)}{2T}.
\label{sus}
\end{eqnarray}
This low temperature behavior should be contrasted with  that of the uniform $S=1/2$ antiferromagnetic Heisenberg chain $C \sim T$ and $\chi \sim \const$ which is less singular than the present Fibonacci case. This is due to the logarithmic divergence of the dynamical exponent in the present case. To check the reliability of the present RSRG scheme, we also applied the same procedure for the Fibonacci XY chain to find
\begin{eqnarray}
\Delta E &\sim &N^{-z}
\end{eqnarray}
with $z=\dfrac{2}{3\ln \phi}\ln\left(\mbox{max}\left(\dfrac{\JA}{\JB},\dfrac{\JB}{\JA}\right)\right)$.  This reproduces the exact result by Kohmoto and coworkers\cite{kkt1} in the limit $\mbox{max}\left(\dfrac{\JA}{\JB},\dfrac{\JB}{\JA}\right) >> 1$. Therefore our RSRG scheme is reliable at least for strong modulation.

The present results appears to be in contradiction with the results of ref. \cite{kh1}, in which the present author carried out the DMRG calculation for the Fibonacci antiferromagnetic Heisenberg chains. In ref. \cite{kh1}, we performed the finite size scaling analysis of the lowest energy gap $\Delta E$ based on the {\it assumption} that it will behave in the same way as the XY chain with relevant aperiodicity, namely as $\Delta E \sim \exp(-cN^{\omega})$. However, the present analysis suggests the different behavior. Although we tried to replot the previous data using the scaling (\ref{true}), the fit turned out to be very poor. The reason of this discrepancy will be understood in the following way.

For finite $\JA/\JB$, the perturbation approximation requires the higher order corrections which modify the recursion equation (\ref{iter}) in the form,
\begin{eqnarray}
\JA^{(n+1)}&=&\frac{\JA^{(n)3}}{\JB^{(n)2}}{f_{\rm A}(\JA^{(n)}/\JB^{(n)})}, \nonumber\\
\JB^{(n+1)}&=&\frac{\JA^{(n)2}}{\JB^{(n)}}{f_{\rm B}(\JA^{(n)}/\JB^{(n)})}.
\label{iterm}
\end{eqnarray}
It should be noted that the correction factors $f_{\rm A}$ and $f_{\rm B}$ depend only on the ratio $\JA^{(n)}/\JB^{(n)}$  and satisfy $f_{\rm A}(0)=1/4, f_{\rm B}(0)=1/2$. This leads to the recursion equation for $\v{X}^{(n)}=(\XA^{(n)},\XB^{(n)})\equiv(\ln \JA^{(n)}, \ln \JB^{(n)})$ as
\begin{eqnarray}
\v{X}^{(n+1)}&=&\left(
\begin{array}{ll}
3 & -2 \\
2 & -1 
\end{array}
\right)  \v{X}^{(n)}+{\v{\mu}(\XA^{(n)}-\XB^{(n)})}
\end{eqnarray}
with $\v{\mu}=(\mu_{\rm A}, \mu_{\rm B})\equiv (\ln f_{\rm A}, \ln f_{\rm B})$. If the function $\v{\mu}(X)$ is approximated by a linear function of $X$ as $\v{\mu}(X)=\v{\gamma}X+\v{\mu}_0$ with $\v{\gamma}=(\gamma_{\rm A}, \gamma_{\rm B})$, we have
\begin{eqnarray}
\v{X^{(n+1)}}&=&M_m \v{X^{(n)}}+\v{\mu}_0,
\label{recur_scml}
\end{eqnarray}
where $M_m$ is a $2 \times 2$ matrix
\begin{eqnarray}
M_m&=&\left(
\begin{array}{ll}
3-\gamma_A & -2+\gamma_A \\
2-\gamma_B & -1+\gamma_B 
\end{array}
\right).  
\end{eqnarray}
One of the eigenvalue of $M_m$ is unity. If another eigenvalue  $\lambda_m(\equiv 1+\gamma_{\rm B}-\gamma_{\rm A})$ is larger than unity, the solution of (\ref{recur_scml}) grows with $n$ as
\begin{eqnarray}
\v{X^{(n)}}&\propto& \lambda_m^n.  
\end{eqnarray}
In this case, both $\ln\JA^{(n)}$ and $\ln \JB^{(n)}$ scale as $\lambda_m^n$. Therefore the lowest energy scale of the chain of length $N$ also scales as,
\begin{eqnarray}
\Delta E&\sim& \exp(-C\lambda_m^{\scriptsize\frac{\ln N}{3\ln \phi}}) 
\sim \exp (-CN^{\scriptsize\frac{\ln \lambda_m}{3\ln \phi}})\nonumber\\
&\sim& \exp (-CN^{\omega})  \ \mbox{with} \ \ \omega\equiv\frac{\ln \lambda_m}{3\ln \phi}
\label{short}
\end{eqnarray}
within appropriate range of system size $N$.  This is the reason why the behavior (\ref{short}) is observed in DMRG calculation for finite systems.  We have numerically diagonalized the Hamiltonian of the clusters BAABABAAB and  BABAABAABAB which reduce to a single A'-bond and B'-bond after decimating B bonds. Using these numerical data, it is verified that the effective value of $\lambda_m$ is larger than unity although it actually depend on $\JA/\JB$.  As the renormalization proceeds, of course, the ratio $\JA/\JB$ decreases and  the true asymptotic behavior (\ref{true}) is reached. More details of this calculation are presented in \cite{kh3}. 

This crossover behavior implies that the extremely low temperature is required to observe the true asymptotic behavior (\ref{spe}) and (\ref{sus}) in weak modulation regime. Instead, the behaviors expected from (\ref{short}), namely $C \sim 1/(\ln T)^{1+1/\omega}$ and  $\chi \sim 1/(T(\ln T)^{1/\omega})$\cite{jh1}, would be observed in the intermediate temperature regime.

In summary, using the RSRG method, we have shown that  the ground state  of the $S=1/2$ Fibonacci Heisenberg chain belongs to a new universality class in which the energy gap scales as $\exp (-\kappa(\ln N)^2)$ where $\kappa$ is a universal constant independent of modulation strength. The low temperature behavior of the magnetic specific heat and magnetic susceptibility is predicted. The relationship to the previous numerical results\cite{kh1}  which appear to contradict with the present calculation is also discussed. The details of the calculation and proof will be reported in a separate paper, which will also includes the discussion of the general XXZ case\cite{kh2} and ground state phase diagram\cite{kh3}.

We have found a new quantum dynamical critical behavior (\ref{true}) which was so far unknown in the field of quantum many body problem. Similar 'singular dynamic scaling' is, however, known since 80's for the classical Ising model on the percolation clusters with Glauber dynamics.\cite{hen} In spite of the geometrical self-similarity common to this classical model and our quantum model, they look very different in many aspects. Although the underlying physics is still unclear, further investigation on this point might lead to a more profound understanding of both systems.

After this work is completed, the preprint by Vieira\cite{vieira} appeared in e-print archive in which some of the present results are derived. In addition, Vieira satisfactorily applied this method to the Heisenberg chains with relevant aperiodicity. Similar approach is also applied to the two dimensional quasicrystal.\cite{jag} We thus expect the RSRG method is widely applicable to various problems in the field of quantum magnetism in quasiperiodic systems.

The author would like to thank Dr. C. L. Henley for drawing his attention to refs. \cite{hen} and for enlightening comments to the eariler version of this paper. This work is supported by a Grant-in-Aid for Scientific Research from the Ministry of Education, Culture, Sports, Science and Technology, Japan.


\begin{thebibliography}{11} 
\bibitem{shecht} D. Shechtman, I. Blech, D. Gratias and J. W. Cahn: Phys. Rev. Lett. {\bf 53}, 1951 (1984).
\bibitem{sato1} T. J. Sato, H. Takakura, A. P. Tsai and K. Shibata: Phys. Rev. Lett. {\bf 81}, 2364 (1998); T. J. Sato, H. Takakura, A. P. Tsai, K. Shibata, K. Ohoyama and K. H. Andersen: Phys. Rev.  {\bf B61}, 476 (2000).
\bibitem{dot1} L. P. Kouwenhoven, F. W. J. Hekking, B. J. van Wees, C. J. P. M. Harmans, C. E. Timmering and C. T. Foxon: Phys. Rev. Lett. {\bf 65}, 361 (1990); R. Ugajin: Physica {\bf E1}, 226 (1997); M. H\"ornquist and T. Ouchterlony: Physica {\bf E3}, 213 (1998).
\bibitem{ledieu} J. Ledieu, J.T. Hoeft, D.E. Reid, J.A. Smerdon, R. D. Diehl, T.A. Lograsso, A. R. Ross and R. McGrath : cond-mat/0307131.

\bibitem{kh1}  K. Hida : J. Phys. Soc. Jpn. {\bf 68}, 3177 (1999). 
\bibitem{kh2}  K. Hida : J. Phys. Soc. Jpn. {\bf 69}, Suppl. A 311  (2000).
 \bibitem{vidal1} J. Vidal, D. Mouhanna and T. Giamarchi: Phys. Rev. Lett. {\bf 83}, 3908 (1999); Phys. Rev. {\bf B65}, 014201  (2002).
\bibitem{acg} M. Arlego, D. C. Cabra, and M. D. Grynberg : Phys. Rev. {\bf B 64}, 134419 (2001); M. Arlego : {\it ibid.}  {\bf 66}, 052419 (2002)
\bibitem{wjh} S. Wessel, A. Jagannathan and S. Haas: Phys. Rev. Lett. 90, 177205 (2003)
\bibitem{jag} A. Jagannathan, Phys. Rev. Lett. {\bf 92}, 047202 (2004) 
\bibitem{kkt1} M. Kohmoto, L. P. Kadanoff and C. Tang: Phys. Rev. Lett {\bf 50}, 1870 (1983); M. Kohmoto and Y. Oono: Phys. Lett {\bf 102A}, 145 (1984); M. Kohmoto, B. Sutherland and C. Tang: Phys. Rev. {B\bf 35},  1020 (1987); H. Hiramoto and M. Kohmoto: Int. J. Mod. Phys. B{\bf 6}, 281 (1992) and references therein.
\bibitem{jh1} J. Hermisson: J. Phys. A: Math. Gen. {\bf 33}, 57 (2000).
\bibitem{taka} For example see M. Takahashi : {\it Thermodynamics of
One-Dimesional Solvable Models}, Cambridge University Press (1999) and references therein.
\bibitem{kore} V. E. Korepin: Commun. Math. Phys. {\bf 110} 157 (1987).
\bibitem{df1} D. S. Fisher: Phys. Rev. B {\bf 50}, 3799 (1994).

\bibitem{kh3}  K. Hida : to be published in J. Phys. Soc. Jpn. 
\bibitem{hen} C. L. Henley : Phys. Rev. Lett. {\bf 54} (1985) 2030; C. K. Harris and R. B. Stinchcombe, {\it ibid.} {\bf 56} (1986) 869; R. Rammal and A. Benoit, {\it ibid.} {\bf 55} (1985) 649.
\bibitem{vieira} A. P. Vieira : cond-mat/0403635.
\end{thebibliography}
\end{document}